\documentstyle[12pt,graphicx,color,subfigure,amsmath,cite]{article}
\def\lsim{\ ^<\llap{$_\sim$}\ }
\def\gsim{\ ^>\llap{$_\sim$}\ }
\def\beq{\begin{equation}}
\def\eeq{\end{equation}}
\def\beqn{\begin{eqnarray}}
\def\eeqn{\end{eqnarray}}

\def\NP{Nucl. Phys.}

\begin{document}
\begin{center}
  { \Large\bf 
Looking for a heavy wino LSP in collider and dark matter experiments
\\}
  \vglue 0.5cm
  Utpal Chattopadhyay$^{(a)}$, Debottam Das$^{(a)}$, 
Partha Konar$^{(b)}$ and D.P.Roy$^{(c)}$
    \vglue 0.2cm
    {\em $^{(a)}$Department of Theoretical Physics, Indian Association 
for the Cultivation of Science, Raja S.C. Mullick Road, Kolkata 700 032, India \\}
    {\em $^{(b)}$
Institut f\"ur Theoretische Physik, Universit\"at Karlsruhe, 
D--76128 Karlsruhe, Germany\\}
 {\em $^{(c)}$Homi Bhabha Centre for Science Education, Tata Institute of
   Fundamental Research,Mumbai-400088, India} \\
%{\em $^{(d)}$Institute for Particle Physics Phenomenology, 
%Ogden Centre for Fundamental Physics, 
%Department of Physics, 
%University of Durham,
%Durham DH1 3LE, United Kingdom
%} 
  \end{center}        
\begin{abstract}
We investigate the phenomenology of a wino LSP as obtained in AMSB and
some string models.  The WMAP constraint on the DM relic density
implies a wino LSP mass of 2.0-2.3 TeV.  We find a viable signature
for such a heavy wino at CLIC, operating at its highest CM energy of
5~TeV.  One also expects a viable monochromatic $\gamma$-ray signal
from its pair-annihilation at the galactic centre at least for cuspy
DM halo profiles.
\end{abstract}

\section{Introduction}
The minimal supersymmetric standard model (MSSM) is the most popular
extension of the  standard model (SM) on account of four attractive
features\cite{Kane:Drees}.  It provides (1) a natural solution to the
hierarchy problem of the SM, (2) a natural (radiative) mechanism for
the electroweak symmetry breaking (EWSB) , (3) a natural candidate for
the cold dark matter (DM) of the universe in the form of the lightest
superparticle (LSP), a prediction that has gained in importance 
in view of recent observations\cite{Clowe:2006eq}, and 
(4) unification of the SM gauge couplings at
the GUT scale.  However it also suffers from two problems:\\ (i)
Little Hierarchy Problem: The LEP limit on the mass of an SM-like
Higgs boson\cite{EdelmanPLB}, 
\beq m_h> \rm 114 \ GeV,
\label{higgs}
\eeq 
requires the average top squark mass to be typically an order
magnitude higher than $M_Z$\cite{HaberNP}. This implies some
fine-tuning of SUSY parameters to obtain the correct value of $M_Z$.\\
(ii) Flavour and CP Violation Problem: The general MSSM makes fairly
large 1-loop contributions to flavour changing neutral current (FCNC)
processes, like $\mu \rightarrow e\gamma$ decay, as well as to CP
violating processes like fermion electric dipole moments (EDM).  The
experimental limits on these FCNC decays require either large scalar
masses, 
\beq 
m_{\phi} \gsim 10 \rm \ TeV
\label{scalar_minlmt}
\eeq
or a near degeneracy among the sfermion masses of different
generations.  Similarly, the experimental limits on the
electron and neutron EDM require either large scalar masses
(as in Eq.\ref{scalar_minlmt}) or unnaturally small CP
violating phases.  It may be added here that while the degeneracy of
sfermion masses can be realised in simple models like minimal
supergravity (mSUGRA) or anomaly mediated SUSY breaking (mAMSB), there
is no simple model ensuring small SUSY phases.

The split SUSY model\cite{splitsusy} tries to solve
the second problem at the cost of aggravating the first by pushing up the 
scalar superparticle masses.  In fact the cost is much more, since this
model assumes the scalar masses to be many orders of magnitude larger 
than the TeV scale.  This means that one has to give up (1) 
the supersymmetric solution to the hierarchy problem of the SM
along with (2) the radiative EWSB mechanism.  One only retains the LSP 
dark matter and the unification of gauge couplings, since the chargino
and the neutralino masses are assumed to remain within a few TeV.  We find the 
cost much too high since the first two features were the original 
motivations for weak scale supersymmetry.

We shall consider instead a more conservative model where the scalar 
superparticle masses are assumed to lie in the range 
\beq
                    m_{\phi} = 10 -100 \rm \ TeV.
\label{scalarlimit}
\eeq
Thus, it solves the second problem at the cost of aggravating the first;
 but without abandoning the supersymmetric solution to the hierarchy problem
or the radiative EWSB mechanism.  Moreover,
we retain the LSP
dark matter as well as gauge coupling unification by assuming the chargino
and neutralino masses to remain within a few TeV.

We shall be primarily interested in the electroweak chargino-neutralino
sector and in particular the lightest neutralino, which we assume to be the 
LSP.  The diagonal elements of the $4\times4$ neutralino mass matrix are
$M_1,M_2,$ and $\pm\mu$, corresponding to the bino $\tilde{B}$, the wino
$\tilde{W}$, and the higgsinos $\tilde H_{1,2}=\tilde H_{u}\pm\tilde H_{d}$,
respectively, while the non-diagonal elements are all $\leq M_{Z}$.  
Now, there are experimental indications from the Higgs mass limit (Eq.~\ref{higgs})
and the $b \rightarrow s \gamma$ decay width that the SUSY masses representing
the above diagonal elements are typically larger than $M_Z$, at least in a 
universal MSSM like the mSUGRA model\cite{ucpn}.  We shall assume
this mass inequality to hold in a more general MSSM.  
Then it implies that the 
neutralino mass eigenstates correspond approximately to the above interaction
eigenstates $\tilde{B}$, $\tilde{W}$ and 
${\tilde H}_{1,2}$ and the LSP constitutes of one of these states. 
An interesting exception to this
rule is provided by the case of a near degeneracy 
between two diagonal 
elements, which results in a large mixing between the corresponding 
interaction eigenstates, as the mixing angle is given by
%\[
$\tan2\theta = 2 \, M_{ij} / |M_{ii}-M_{jj}|$ .
%\]
In particular the LSP can be a mixed bino-higgsino, bino-wino 
or wino-higgsino state. Such mixed LSP cases have been 
investigated in Ref.\cite{mixed_lsp,mixed_lsp_ark}, and
named ``well-tempered'' neutralino in Ref.\cite{mixed_lsp_ark}. 

Leaving aside such an accidental degeneracy between the two lightest mass 
eigenvalues, one expects the LSP to be approximated by one of 
the interaction states - bino, wino or higgsino. 
The bino carries no
gauge charge and hence does not couple to gauge bosons.  Thus it can only
pair-annihilate via sfermion exchange.  But the current experimental
lower limits on the sfermion masses\cite{EdelmanPLB} imply a low
annihilation rate, resulting in an overabundance of dark matter over
most of the MSSM parameter space.  Only in special regions like stau
co-annihilation ($M_{1} \simeq m_{\tilde \tau{1}}$) or resonant
pair annihilation ($2M_{1} \simeq m_{A}$) can one get a cosmologically
acceptable DM relic density.  But neither of these regions extend to the
scalar mass range of Eq.(\ref{scalarlimit}).  Even the so called focus
point region, which corresponds to a ``well-tempered'' bino-higgsino
LSP, does not reach the scalar mass range of Eq.\ref{scalarlimit} in the
universal MSSM\cite{Feng,Chatto-etal}, although, 
in a generic and  unconstrained MSSM,  it can obtain 
the correct DM relic density for very large scalar masses\cite{mixed_lsp_ark}.
  In contrast, 
the higgsino and wino carry isospins
$1/2$ and $1$ respectively. Hence they can pair annihilate to
\beqn
\tilde{H}\tilde{H} \rightarrow WW({\bar f} f) , \  
\tilde{W}\tilde{W} \rightarrow WW({\bar f} f),  
\label{hig-win-ann}
\eeqn
by their gauge couplings to W boson.
Consequently, the annihilation
rate and the resulting DM relic density is controlled mainly by the
higgsino (wino) LSP mass. It has only a marginal dependence  
on the sfermion mass\cite{mixed_lsp_ark} and it is 
practically independent of the other SUSY parameters.  The
current WMAP result on the DM relic density alongwith 
the $2\sigma$
error bar\cite{spergel} is
\beq
\Omega_{\tilde \chi}h^2=0.104 ^{+0.015} _ {-0.019}
\label{relicdensity}
\eeq  
where $h=0.73\pm0.03$ is the Hubble constant in units of 
$100 \ \rm Km \ \rm s^{-1}
\ \rm Mpc^{-1}$\cite{spergel} and $\Omega_{\tilde \chi}$ 
is the DM relic density in units of the critical density.  
This corresponds to a higgsino (wino) LSP mass
of about $1 \ (2)$ TeV, where the larger wino mass is due to its larger
gauge coupling.  In ref.\cite{Chatto-etal} we investigated the
phenomenology of higgsino LSP in collider and dark matter experiments.  
The present work is devoted to a similar investigation for the wino LSP.

In the next section,
 we discuss the wino LSP models and estimate the wino
mass band compatible with the WMAP relic density range of Eq.\ref{relicdensity}
.  In the two following sections we investigate the prospects of detecting
such a heavy wino LSP in future collider and DM search experiments 
respectively.  Finally we shall conclude with a summary of our results.
 
\section{Wino LSP in AMSB and string models:}
\label{winoLSP}
A universal gaugino mass at the GUT scale leads to the 
weak scale wino being
always heavier than the bino, 
since the gaugino masses evolve like the 
corresponding
gauge couplings.  Hence, 
the wino LSP scenario can not be realised in the 
universal MSSM.  The most popular SUSY model for a wino LSP is the anomaly 
mediated supersymmetry breaking (AMSB) model\cite{randall,giudice}
wherein 
the gaugino and scalar masses arise 
from supergravity breaking in the 
hidden sector via super-Weyl anomaly contributions\cite{gherghetta}, 
namely, 
\beq
M_\lambda={\frac{\beta_g}{g}} \, m_{3/2}
\label{gauginoreln}
\eeq
\beq
m^2_\phi=-\frac{1}{4} \, \left({\frac{\partial\gamma}{\partial g}}\beta_g+
{\frac{\partial\gamma}{\partial y}}\beta_y\right) \, m^2_{3/2}
\label{scalar_gaugino}
\eeq
\beq
A_y=-{\frac{\beta_y}{y}} \, m_{3/2}.
\label{A-gaugino}
\eeq
Here $m_{3/2}$ is the gravitino mass, $\beta_{g}$ and $\beta_y$ are the 
$\beta$ functions for gauge and Yukawa couplings, and 
$\gamma = {\partial ln Z}/
{\partial ln \mu}$, where $Z$ is the wave function renormalization constant.
The GUT scale gaugino masses (\ref{gauginoreln}) are 
thus non-universal, with 
\beqn
M_1={\frac{33}{5}} \, {\frac{g_1^2}{16\pi^2}} \, m_{3/2} \ , \qquad
M_2={\frac{g_2^2} {16\pi^2}} \, m_{3/2} \ , \qquad 
M_3=-3 \, {\frac{g_3^2}{16\pi^2}} \, m_{3/2}
\label{gauginoNU}
\eeqn
at the one loop level.  Evolving down to the weak scale gives 
\beq
M_1 : M_2 :|M_3| \simeq 2.8 : 1 : 7.1
\label{gaugino-ratio}
\eeq
including the two loop corrections.  Unfortunately, 
evolving the scalar masses
of Eq.(\ref{scalar_gaugino}) down to the weak scale gives negative 
mass-square values for sleptons.  
In the minimal version of the model (mAMSB) this is remedied by 
adding a common 
parameter $m_0^2$ to the right hand side of Eq.(\ref{scalar_gaugino})
for all the scalars in the theory.  
This model has
been widely studied because of its economy of parameters, {\em i.e.,} 
\beq
m_{3/2}, \ m_0,\ \tan\beta, \, {\rm sgn}(\mu)
\label{GUTinput}
\eeq
$\mu^2$ being fixed by the radiative EWSB condition.  We shall come
back to this model below.

It should be noted here that the anomaly mediated contributions of 
Eqs.(\ref{gauginoreln},\ref{scalar_gaugino},\ref{A-gaugino})
are present  in all supergravity models.  But, in general,
 one can also have
tree level SUSY breaking contributions to the gaugino and scalar masses
arising from possible dimension five and six terms in the effective 
Lagrangian, namely
\beq
M_{\lambda} \in {\frac{F_S}{M_{pl}}} \, \lambda \, \lambda
\label{lambdafield}
\eeq
and
\beq
m^2_\phi \in {\frac{F_S^{\dagger} \, F_S}{M_{pl}^2}} \, {\phi^\star}\phi
\label{phifield}
\eeq
where $F_S$ is the vev of the $F$ component of a chiral superfield
$S$ responsible for SUSY breaking.  If present, these tree
level contributions are expected to overwhelm the anomaly mediated
contributions of Eqs.(\ref{gauginoreln}\& \ref{scalar_gaugino}).  
The AMSB scenario assumes 
the SUSY breaking superfield
to be carrying a non-zero gauge charge, so that the gaugino mass term
(Eq.\ref{lambdafield}) is eliminated by gauge symmetry.  
However, such symmetry 
considerations can not eliminate the tree level 
scalar mass term (Eq.\ref{phifield}).  So in this
case the scalar mass is expected to be typically larger than the 
gaugino mass by a loop factor, namely
\beq
m_\phi  \sim 100 \, M_{\lambda} \ .
\label{philambda}
\eeq

  This was the case in the AMSB model of Ref.\cite{giudice} 
which has been 
revived in 
Ref.\cite{wellsPRD}.  On the other hand, the
 mAMSB model \cite{randall,gherghetta} 
assumes the SUSY-breaking
hidden sector and the visible sector to reside on two different branes,
separated by a large distance in a higher dimensional space, so that the
tree level scalar mass term (Eq.\ref{phifield}) 
is suppressed by geometric 
considerations.  We shall consider both the possibilities here.

Note also that one can get a AMSB like scenario in the string theory,
where the tree-level SUSY breaking masses can only come from the
dilaton field, while they receive only loop contributions from
moduli fields.  In fact, 
 such a scenario was already suggested in
Ref.\cite{Ibanez-munoz} before the AMSB model by assuming that SUSY breaking is
dominated by a modulus field.  It contributes to the gaugino mass
$M_{\lambda}$ as well as the squared scalar mass $m_{\phi}^2$ at the
1-loop level.  The resulting hierarchy of gaugino masses is very
similar to that of the AMSB (\ref{gauginoNU}).  
On the other hand, the scalar mass
here is expected to be typically larger than the gaugino mass by the
square root of the loop factor, {\em i.e.}
\beq
m_\phi \sim 10 M_\lambda .
\label{philambda_2}
\eeq
Note that the range (\ref{scalarlimit}) is roughly compatible with both 
Eqs.(\ref{philambda} \& \ref {philambda_2}).

As mentioned earlier, we expect the DM relic density to be determined 
by the wino mass $M_2$, practically independent of the other model 
parameters. To check this,
 we have computed the DM relic density, using the
{\it Micromegas} code\cite{micromegas}, 
as a function of $M_2$ with the 
corresponding $M_1$
and $M_3$ determined 
from the AMSB relation (Eq.\ref{gaugino-ratio}).  
With the above gaugino mass relation, the wino mass values for 
the three values (lower, central and upper) of the WMAP relic density
of Eq.(\ref{relicdensity}) 
corresponding to a common sfermion mass of $10$~TeV are 
\begin{equation}
(M_2,\Omega_{\tilde \chi}h^2): \qquad 
(1.91~\rm TeV,0.084 ), \, (2.10~\rm TeV,0.104), \, 
(2.23~\rm TeV,0.119).
\label{winolowhigh}
\end{equation}
The other chosen SUSY parameters were $\mu = 9$ TeV and $\tan\beta = 10$.
In other words, 
the wino mass range corresponding to the $\pm 2 \sigma$
range of the WMAP relic density is 
\beq
M_2 \simeq 1.9-2.2 \rm \ TeV.
\label{m2limit}
\eeq
We have confirmed
these results
using the DARKSUSY code\cite{darksusy} and cross checked 
them
with the results obtained by Profumo\cite{profumo}.  
We have also checked that changing the sfermion mass from 
10 to 100 TeV changes the wino mass upper 
limit of (\ref{winolowhigh}) from 2.23 to 2.37 TeV, due to the 
vanishing of the 
small but negatively interfering sfermion contribution,  
while it has practically no dependence on $\mu$ and $\tan\beta$.

We have also done a more detailed analysis using the mAMSB model. 
We have estimated the weak scale superparticle masses from the
GUT scale input parameters (\ref{GUTinput}) via two loop RGE using the SUSPECT
code\cite{suspect}. The superparticle masses were then used in the
{\it Micromegas} to compute the DM relic density.  
\begin{figure}[!h]
\centering
\includegraphics[width=0.7\textwidth,height=0.5\textwidth]{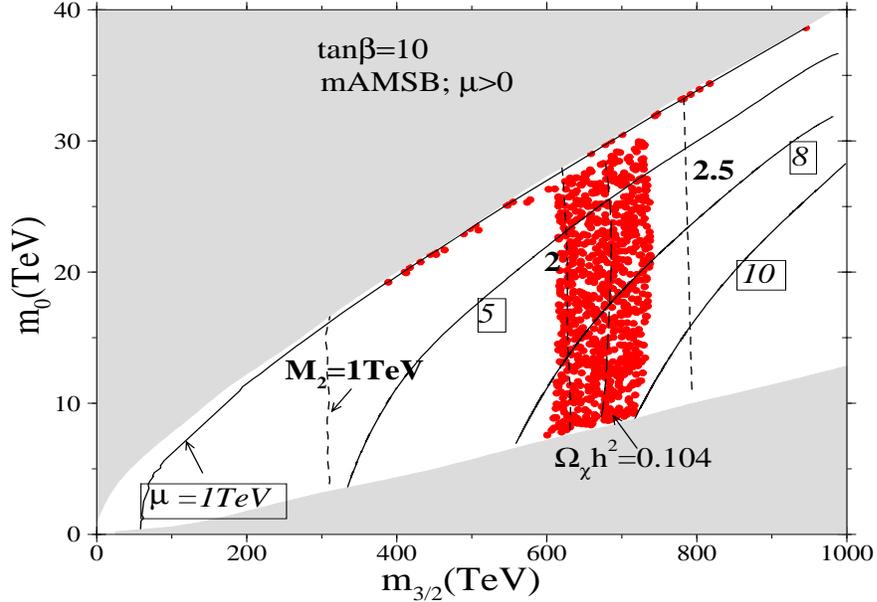}
\caption{Allowed and disallowed regions of the mAMSB parameter space with 
contours shown for $\mu$ (solid lines) and $M_2$ (dashed lines). 
The WMAP satisfied neutralino 
relic density 
zone is shown by red/dark dots which is concentrated in the region of 
$M_2=2.0-2.3 \rm \ TeV$. A contour for 
$\Omega_{\tilde \chi}h^2=0.104$~(central value) is also shown. Isolated 
dots near the REWSB boundary correspond to Higgsino dominated LSP regions. 
The upper shaded (gray) region is disallowed via $\mu^2 < 0$,{\em i.e.}  
the first REWSB constraint.  
The lower shaded (gray) region is disallowed as it contains   
a tachyonic slepton zone (for smaller $m_{3/2}$) and a tachyonic 
pseudoscalar Higgs zone (for larger $m_{3/2}$), {\em i.e.} from the second REWSB constraint.}
\label{m32vsm0}
\end{figure}

Fig.~\ref{m32vsm0} shows the allowed parameter space
in the $m_{3/2}-m_0$ plane for $\tan\beta = 10$ and positive $\mu$.
The upper disallowed region corresponds to $\mu^2 < 0,$ {\em i.e.} no
radiative EWSB, while the lower disallowed region corresponds to
either a tachyonic slepton (for small 
$m_{3/2}$) or a tachyonic 
pseudoscalar Higgs (larger $m_{3/2}$).  
It also shows contours of fixed $\mu$ and fixed $M_2$ in 
the allowed region of parameter space.  The dotted band shows the 
part compatible with the WMAP relic density range of eq.(\ref{relicdensity}),
that is the wino mass band 
\beq
M_2 = 1.9-2.3 \rm \ TeV.
\label{m2-amsblmt}
\eeq
Note also a thin dotted band around the $\mu=1$ contour, representing a 
higgsino LSP.  Thus, 
a WMAP satisfying higgsino LSP of mass of $\sim 1$~TeV 
is realised in mAMSB as well as in the mSUGRA model\cite{Chatto-etal}.  The
 lower end of wino mass corresponds to slepton and 
squark masses in the range of
$9-14$ TeV, while the upper end corresponds to these masses in the range of
$28-30$ TeV.  All these results are very 
insensitive to changes in $\tan\beta$ and
$sign(\mu)$.  It should be noted 
here that the region below the WMAP compatible
wino mass band of Eq.\ref{m2-amsblmt}
 corresponds to an under-abundance of the DM 
relic density
in the standard cosmological model.  This region may be allowed if there are 
alternative DM candidates, or, more interestingly,
 if there are non-standard
cosmological mechanisms for enhancing the relic density of the wino DM.  In
fact, both thermal and non-thermal mechanisms 
for enhancing the wino relic
density have been suggested in the literature\cite{salatiPLB,
Nihei:2005qx,murakami-co}. 
 In the first case, 
the presence of a quintessence field leads to faster Hubble expansion and hence an earlier
freeze-out, resulting in a higher thermal relic density 
\cite{salatiPLB}.  In the second case, 
late decay of the gravitino enhances the wino relic density in the AMSB model
\cite{gherghetta,murakami-co}.  Therefore,
 while investigating the wino 
LSP signal in collider and
dark matter search experiments in the next two sections we shall cover wino 
masses below the band of Eq.\ref{m2-amsblmt} as well.

\section{Wino LSP search in collider experiments}
\label{Winocollider}
Evidently, the wino mass band of Fig.~\ref{m32vsm0} 
is way above the discovery reach of
LHC in the AMSB model\cite{baer-tataPLB}.  In fact, 
the total wino pair 
production cross-section at LHC (${\tilde{W}^\pm}{\tilde{W}^0}+{\tilde{W}^+}{\tilde{W}^-}$) is only 
$\sim 10^{-2} \rm \ fb$\cite{mixed_lsp}, corresponding to $1$ event per year even at the high 
luminosity run of LHC.  Moreover, the mass degeneracy of ${\tilde{W}^\pm}$ and
$\tilde{W}^0$ implies that the only visible objects in the final state will be
1 or 2 soft pions from the 
\beq
{\tilde{W}^\pm} \rightarrow {\pi^{\pm}}{\tilde{W}^0}
\label{wino-pion}
\eeq
decay.  It will be impossible to identify such events without an effective 
tag at the LHC.

The most promising machine for detecting a wino LSP of mass up to the $2 \rm \ TeV$ 
range is the proposed ${e^+}{e^-}$ linear collider CLIC, operating at its 
highest energy of 5~TeV\cite{CLIC}.  We shall follow the strategy of 
Ref.\cite{chen-drees} in 
estimating the signal and the background.  The same strategy has been 
followed by
the LEP experiments in setting mass limits on a wino LSP\cite{EdelmanPLB}; 
 in particular,
the OPAL experiment has used it to set a mass limit of $90$~GeV\cite{OPAL}.  
The pair
production of charged wino is tagged by a hard photon from initial state 
radiation (ISR), {\em i.e.} 
\beq
e^+e^- \rightarrow \gamma {\tilde{W}^+}{\tilde{W}^-}.
\label{ewgamma}
\eeq
The photon is required to have an angle 
\beq
170^\circ > \theta_\gamma > 10^{\circ}
  \label{photon-angle}
\eeq
relative to the beam axis. 
Moreover, it is required to satisfy 
\beqn
E_T^{\gamma} > E_T^{\gamma_{min}} = \sqrt{s} \
\frac{\sin\theta_{min}}{1+\sin\theta_{min}} = 100 \rm \ GeV,
\label{cond-egamma}
\eeqn which vetoes the radiative Bhaba background $e^+e^- \rightarrow
\gamma e^+e^-$, by kinematically forcing one of the energetic $e^\pm$
to emerge at an angle $>\theta_{min}$.  At the maximum CLIC energy of
$\sqrt{s}=5 \rm \ TeV$, the above $E_T^{\gamma_{min}}$ of $100$~GeV
implies $\theta_{min} \simeq 1.2^\circ$.  The OPAL detector has
instrumentation down to $\theta_{min}=2^\circ$, while it seems
feasible to extend it down to $1^\circ$ at future linear
colliders\cite{chen-drees}.  We shall also impose the recoil mass cut
\beq 
M_{rec}={\sqrt{s}} \, 
    \left( 1-{\frac{2E^\gamma}{\sqrt{s}}} \right)^{\frac{1}{2}}>
2 \, m_{\tilde \chi},
\label{rec-mass}
\eeq
where $m_{\tilde \chi}$ represents the LSP mass ($=M_2$ for wino).  This is automatically satisfied 
by the signal (Eq.\ref{ewgamma}).  

In calculating the cross-section of Eq.\ref{ewgamma} 
we have included ISR effects by convoluting the hard $2\rightarrow3$ 
cross-section with the electron
distribution function as described in Ref.\cite{drees-godbole}.  
Although a negatively interfering $t-$channel 
$\tilde{\nu}_e$ exchange contribution
reduces the above cross-section for smaller sneutrino masses, the decrease 
is $\lsim 15$\% for our region of interest (Eq.\ref{scalarlimit}) 
where sfermion mass is higher than $10 \rm \ TeV$.
Hence, we have neglected 
the sneutrino exchange contribution.  

If we can not identify the $\tilde{W}^\pm \rightarrow \tilde{W}^0$ decay products then the
main background is 
\beq
e^+e^- \rightarrow \gamma\nu\bar{\nu}.
\label{egamm-neut}
\eeq
\begin{figure}[!h]
%\vspace{-4cm}
\centering
\includegraphics[width=0.7\textwidth]{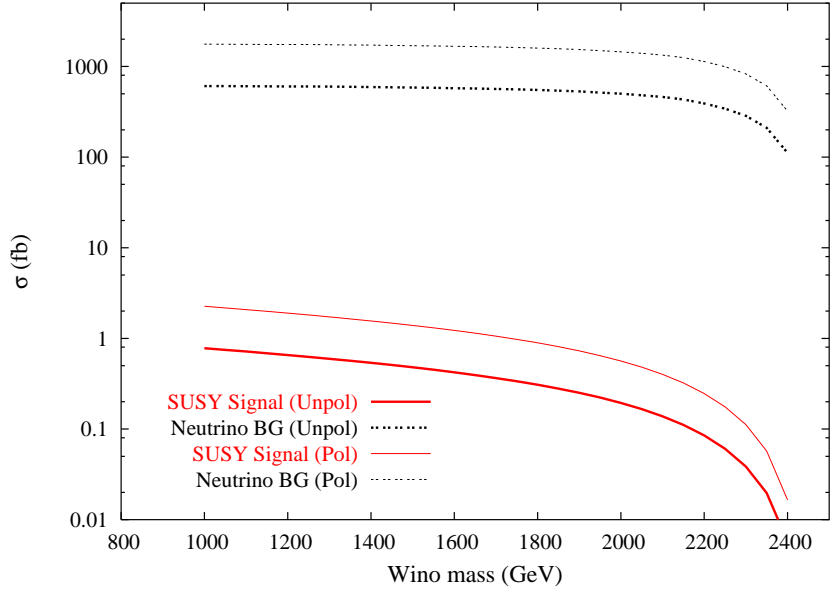}
\caption{
Cross-sections for the wino signal (solid) and the 
neutrino background (dashed) 
at CLIC ( $\sqrt{s} = 5$~TeV) with both unpolarised (thick
lines) and polarised (thin lines) $e^-$ and $e^+$ beams. 
Initial state radiation is included.
}
\label{LspVsSigma}
\end{figure}
Fig.\ref{LspVsSigma} shows the signal and background cross-sections as
function of the LSP mass.  The latter is seen to be larger by
a factor of $\sim 1000$ owing to the large contribution from
the $t-$channel $W$-exchange contribution to $\gamma\nu_e\bar\nu_e$
production.  In the case of higgsino LSP signal,
 this background
could be suppressed by using 
right (left) polarised $e^- \ (e^+)$
beam\cite{Chatto-etal}.  Unfortunately, this does not help here since
the wino pair production signal (\ref{ewgamma}) can arise 
only from the
left (right) polarised $e^- \ (e^+)$ collision.  On the other hand,
one
can enhance the signal cross-section along with the background by
using left (right) polarised $e^- \ (e^+)$ beam.  We have estimated
the signal and background cross-sections for beam
polarisations similar to that 
envisaged for ILC\cite{LHC/ILC}, {\em i.e.}
\beqn
P_{e^-}=-0.8 \ ({\rm mostly \ left \ handed}),
P_{e^+}=0.6 \ ({\rm mostly \ right \ handed}). 
\eeqn
It is easy to check that it corresponds to the following fractional luminosities,
\beq
e_L^-e_R^+:e_R^-e_L^+:e_R^-e_R^+:e_L^-e_L^+ = 0.72:0.02:0.08:0.18
\label{eler-ratio}
\eeq 
while each was 0.25 in the unpolarised case.  It results in increase of both the signal
and the background cross-sections by a factor of $0.72/0.25 \simeq 3$, as shown in Fig.~\ref{LspVsSigma}.

Evidently,
 it is essential to identify the $\tilde{W}^\pm \rightarrow \tilde{W}^0$ decay
products for extracting the signal (Eq.\ref{ewgamma}) 
from the much larger background 
(Eq.\ref{egamm-neut}).  Indeed it is possible to identify 
these decay products unambiguously unlike those for
the higgsino case\cite{Chatto-etal}, thanks to a robust prediction 
for the $\tilde{W^\pm}$
and $\tilde{W^0}$ mass difference $\delta m$\cite{gherghetta,hisano}, which
largely arises from radiative corrections. The gauge boson 
loops give\cite{pierce-bagger}
\beq 
\delta m= {\frac{\alpha \, 
M_W}{2 \, (1+\cos\theta_W)}} \; 
\left[ 1-{\frac{3}{8\cos\theta_W}}{\frac{M_W^2}{M_2^2}}\right] 
\simeq 165\rm \ MeV
\label{delta-m}
\eeq
with the approximate equality holding only 
for $\mu \gg M_2 \gg M_W$. For $M_2 \sim 2$ TeV and $\mu > M_2$ 
, the
 region of our interest, it
gives
\beq
\delta m= 165-190 \rm \ MeV.
\label{numercal-dm}
\eeq 
The tree-level contribution to $\delta m$ is 
$\simeq \tan^2\theta_W \, \sin^2{2\beta}  \, 
M_W^4 / (M_1 \, \mu^2) < 1$ MeV.  
Similarly, the sfermion exchange loop contribution to $\delta m$ is 
${\cal O}(M_W^4 / m_\phi^3) < 1$ MeV.

The mass difference of Eq.(\ref{numercal-dm}) implies ${\tilde{W}^\pm}
\rightarrow {\pi^{\pm}}{\tilde{W}^0}$ to be the dominant decay mode
with a range $c\tau = 3-7$ cm, nearly independent of the wino
mass\cite{chen-drees}.  Moreover, as was pointed out
in\cite{chen-drees}, the SLD vertex detector has its innermost
layer at $2.5$ cm from the beam and this gap is proposed to be reduced
to $2$ cm or even less at the future linear colliders.  Thus, 
it should
be possible to observe the tracks of ${\tilde W}^\pm$ as two heavily ionising
particles along with their decay $\pi^\pm$ tracks in vertex detector.
Moreover for the momentum of the decay pion, $p_\pi \sim \sqrt{\delta
m^2 - m_{\pi}^2} \sim 87-128$ MeV, one expects the impact parameter
resolution to be better than $0.3$ mm.  Thus both the decay pions have
impact parameters of $\gsim 100\sigma$, which should be easily
measurable.  These should enable us to distinguish the signal
(Eq.\ref{ewgamma}) unambiguously from the background 
(Eq.\ref{egamm-neut})
even in the presence of the beamstrahlung pions\cite{drees-PRL}.
Therefore we expect the viability of the signal to be determined
primarily by the number of signal events.

We see from Fig.~\ref{LspVsSigma} that, with the proposed luminosity of $1000 fb^{-1}$ at CLIC\cite{CLIC}, one
expects 600 (200) to 120 (40) events with polarised (unpolarised) beams for the WMAP satisfying
wino mass range of $2.0$ to $2.3$~TeV (Fig.\ref{m32vsm0}).  
It should be noted here that the search can be extended to wino
mass of 2.4 (2.5) TeV with a proportionate increase of the beam energy
by 5 (10)\%.

\begin{figure}[!h]
%\vspace{-4cm}
\centering
\includegraphics[width=0.7\textwidth]{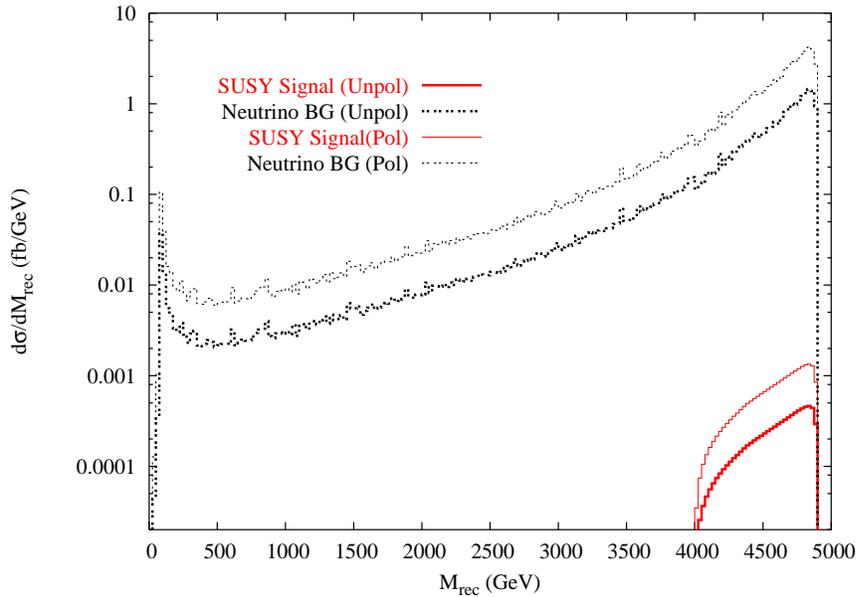}
\caption{
Recoil mass distributions of a 2 TeV wino
signal and the neutrino background at CLIC ( $\sqrt{s} = 5$~TeV) with
both unpolarised (thick lines) and polarised (thin lines) $e^-$ and $e^+$
beams. Initial state radiation is included.
}
\label{recoilfig}
\end{figure}
Fig.~\ref{recoilfig} shows the recoil mass distribution of a $2$~TeV wino LSP signal along with the background events.  While the recoil mass distribution of 
the background (\ref{egamm-neut}) stretches all the way from $M_Z$ up to the kinematic limit, the signal
shows a characteristic threshold at $2m_{\tilde \chi}$. This 
will help confirm the signal
as well as to measure the LSP mass $m_{\tilde \chi}$.  

\section{Wino LSP search in DM experiments}
\label{winoDMsearch}
The wino LSP signal is too small to be observed in direct dark matter
search experiments.  The reason is that this signal comes from
spin-independent ${\tilde \chi}p$ scattering, which is dominated by
Higgs boson($h,H$) exchange.  Since the Higgs coupling to the LSP pair
is proportional to the product of their higgsino and gaugino
components, it is vanishingly small for the wino LSP. 
However, there can be significant contribution to the spin-independent
scattering cross-section from one loop diagrams as shown in 
Ref.\cite{Hisano:2004pv}.
The signal is
further suppressed by the large LSP mass.  Likewise, the neutrino
signal coming from the pair annihilation of wino LSP in the solar core
is vanishingly small.  This is because the solar capture rate of the
LSP is controlled by the spin-dependent ${\tilde \chi}p$ scattering
cross-sections via $Z$ boson; and the $Z$ coupling to $\tilde \chi$
pair is proportional to the square of its higgsino component.

A very promising wino DM signal is expected to come from
$\gamma$-rays produced by its pair annihilation at the galactic
centre.  The largest signal comes from the tree-level annihilation
process (\ref{hig-win-ann}) into $WW$, followed by the decay 
of the $W$ into
$\gamma$-rays via neutral pions\cite{mixed_lsp}. Unfortunately, the
continuous energy spectrum of the resulting $\gamma$-rays suffers from
a large background from the cosmic-ray pions.  We consider instead the
monochromatic $\gamma$-ray signal coming from the annihilation process
\beq \tilde W \tilde W \rightarrow \gamma \gamma , \gamma Z
\label{DM-annih}
\eeq
via ${\tilde{W}^\pm}{W^\mp}$ loops\cite{bergstrom-ullio}. 
For the resulting cross-sections,
\beq
v\sigma_{\gamma \gamma} \sim v\sigma_{\gamma Z} \sim 10^{-27} \rm \ cm^3 \rm \ s^{-1},
\label{DM-xsection}
\eeq
where $v$ is the velocity of the DM particles in their cms frame. The resulting $\gamma$-ray flux coming from an angle $\psi$ relative to the galactic centre can be written as
\cite{bergstrom-ullio} 
\beq
\Phi_{\gamma}(\psi)=1.87 \times 10^{-13} {\frac{N_{\gamma}v\sigma}{10^{-27} cm^{3} s^{-1}}} ({
\frac{1 \rm \ TeV}{m_{\tilde \chi}}})^2 J(\psi) \rm \ cm^{-2}\rm \ s^{-1} \rm \ sr^{-1}
\label{phipsi}
\eeq
where $N_\gamma=2 \ (1)$ for the $\gamma \gamma \ (\gamma Z) $ production; and
\beq
%J(\psi)=
J(\psi) =\frac{\int_{line \ of \ sight} \rho^2(\ell) d\ell(\psi)}{\left[ (0.3 
\rm \ GeV/{\rm cm}^3)^2 \cdot 8.5 \ {\rm kpc} \right]}
%Gev/{cm}^3)^2 \cdot 8.5 \ {kpc} \right]}
\label{line-integral}
\eeq 
is the line integral of the squared DM energy density scaled by its local value in our
neighbourhood  and our distance from the galactic centre.

Several Atmospheric Cerenkov Telescopes (ACT) have started recording
TeV scale $\gamma$-rays from the galactic center e.g. 
HESS and CANGAROO in the
southern hemisphere and MAGIC and WHIPPLE in the north.  One generally
expects concentration of DM in the galactic centre; but its magnitude
has a large uncertainty depending on the assumed profile of DM halo
density distribution\cite{navarro,diemand,burkert}.  The cuspy NFW
profile\cite{navarro} corresponds to
\beq
\langle J(0)\rangle_{\Delta \Omega = {10}^{-3}} \simeq 1000,
\label{J0}
\eeq
which represents the DM flux in the direction of the galactic centre
averaged over the typical ACT aperture of $\Omega=10^{-3}$ sr.
Extreme distributions, like the spiked profile\cite{diemand} and core
profile\cite{burkert}, correspond respectively to increase and
decrease of this flux by a factor of $10^{3}$.  We have computed the
$\gamma$-ray line signal (\ref{phipsi}) in the mAMSB model for the NFW
profile and an aperture $\Delta \Omega=10^{-3}$ sr using the DARKSUSY
code\cite{darksusy}.
\begin{figure}[!h]
\centering
\includegraphics[width=0.7\textwidth,height=0.5\textwidth]{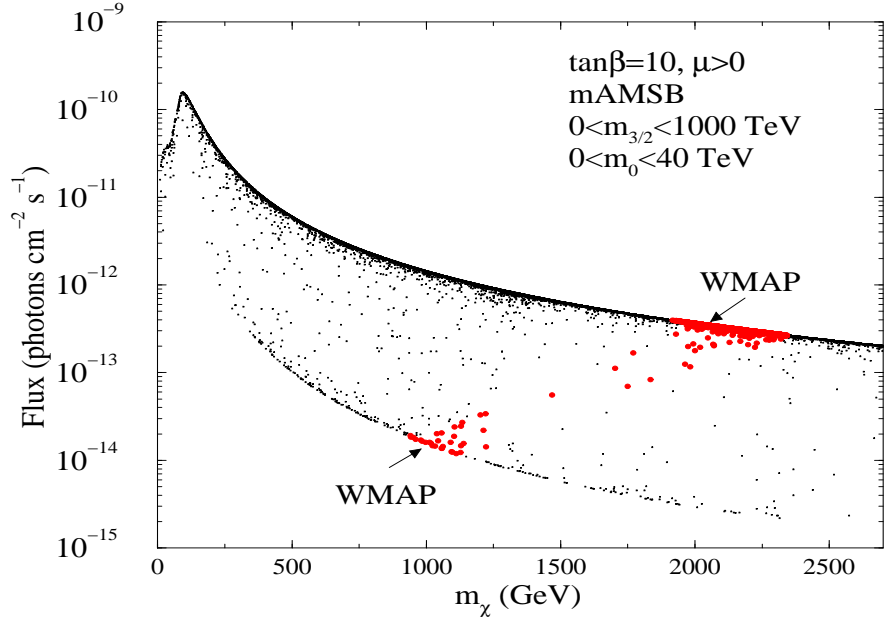}
\caption{Monochromatic $\gamma$-ray flux from LSP pair annihilation at the 
galactic center for the NFW profile of DM halo distribution with an aperture 
of $\Delta \Omega=10^{-3}$~sr for mAMSB model for varying LSP mass. The red (darker) points correspond to the WMAP relic density satisfied values of 
eq.\ref{relicdensity}}
\label{photonflux}
\end{figure}
Fig.~\ref{photonflux} shows the resulting signal against the LSP mass, where we have added the
$\gamma \gamma$ and $\gamma Z$ contributions, since they give identical photon energy
($=m_{\tilde \chi}$) within the experimental resolution.  The points satisfying WMAP relic density
are shown as bold dots.  One clearly sees a wino LSP of $2$-$2.3$ TeV mass predicting a line
$\gamma$-ray flux of $\sim 10^{-13} \rm \ {cm}^{-2} \rm \ s^{-1}$ along with a higgsino LSP of $1$  TeV mass predicting a flux of 
$\sim 10^{-14} \rm \ {cm}^{-2} \rm \ s^{-1}$.  Both these are in agreement with the results
of reference\cite{hisano} and\cite{thomas,Mambrini:2005vk}\footnote{It has been pointed 
out recently\cite{Bergstrom:2005ss} 
that tree-level higher order processes, in particular $\chi \chi \rightarrow W^+ W^- \gamma$, can increase the flux of photons with $E_\gamma \simeq 
m_\chi$ by up to a factor of $2$. }.  They are within the detection range of the above-mentioned
ACT experiments.  In particular, the wino signal has the advantage of an order of magnitude 
higher flux compared to the higgsino LSP.  Furthermore, 
its higher mass implies an order of magnitude lower background from
cosmic-ray proton and electron showers, as shown in Ref.\cite{thomas}.  It is
further shown in \cite{thomas} that one expects to see a $5 \sigma$ wino signal over this background at these ACT 
experiments for a NFW (or cuspier) profile.

However, it should be noted here that the HESS experiment has
reported TeV photons coming from the direction of the galactic centre
with an energy spectrum, which is unlike that expected from a TeV
scale $\gamma$-ray line\cite{HESS}.  Instead, it shows a
power law decrease with energy, which is similar to that of other
"cosmic accelerators", notably the supernova remnants (SNR).  Besides
this, it is not clear whether this signal is coming right
from our galactic centre (defined as the location of the super-massive
black hole Sagittarius $A^\star$), or from a nearby SNR lying within
the angular resolution of HESS.  In particular, the SNR
Sagittarius $A$ east, lying a few parsecs away from Sagittarius
$A^\star$ has been suggested to be the culprit\cite{grasso}.  Given
the modest energy resolution of the present ACT experiments ($\sim 15
\%$), it may be more difficult to extract a line $\gamma$ ray signal
at the $5 \sigma$ level in the presence of this $\gamma$ ray
background.  Therefore it is imperative to improve the energy and
angular resolution of these experiments to suppress this background;
and also to look for other possible clumps of DM in galactic
halo\cite{Hopper}. We point out here that in our discussion we 
have considered the
neutrino and gamma ray signals to probe the known sources of DM concentration
like the solar and the galactic cores because of their directionality. 
On the other hand the DM annihilation
in the galactic halo can also be 
probed via the positron and the anti-proton channels as discussed in 
Refs.\cite{mixed_lsp,Hisano:2005ec}. Indirect detection of wino 
as a dark matter candidate satisfying WMAP data via cosmic ray 
positron and anti-proton fluxes particularly becomes more 
interesting because of non-perturbative enhancement of 
cross-sections\cite{Hisano:2005ec}. 

We shall conclude with a brief discussion of the non-perturbative
contribution to the annihilation cross-sections of TeV 
scale wino LSP and its impact on our results. The non
perturbative contributions coming from the s-channel bound states, 
as calculated in \cite{hisano,Hisano:2003ec} via an effective potential, 
leads to a large enhancement of the wino pair annihilation
cross-section into WW and $\gamma$-$\gamma$ channels. In a recent work 
of non-perturbative calculation\cite{Hisano:2006nn} 
the above authors have computed the velocity averaged wino pair
annhilation cross-section at the freeze-out temperature
and the resulting wino relic density.
They show a $\sim$50{\%} reduction of the thermal abundance,
with respect to the perturbative value corresponding to 
$\sim$600~GeV or a 
$\sim$25{\%} increase in the wino mass satisfying the 
WMAP relic density. The
collider signal for such a wino LSP will require a
proportionate $\sim$25{\%} of the CLIC beam energy.
On the other hand the line $\gamma$-ray signal will be 
larger that that shown in Fig.~\ref{photonflux}. The
corresponding enhancement of the continuum gamma ray 
signal has been discussed in Ref.\cite{hisano}, while
the positron and the anti-proton signals have been 
discussed in Ref.\cite{Hisano:2005ec}. 

\section{Summary}
\label{secConclusion}
1)  We study the phenomenology of a wino LSP obtaining in the 
AMSB and some string models.

\noindent
2)  The WMAP constraint on the DM relic density implies a heavy 
wino LSP mass of $2.0-2.3$  
 TeV in the standard cosmology. But one can also have wino LSP 
mass $<2$ TeV assuming 
nonstandard cosmological mechanisms for enhancing the DM relic density.

\noindent
3)  We find a viable wino LSP signal all the way upto $2.3$ TeV at the proposed $e^+e^-$
linear collider (CLIC), operating at its highest CM energy of $5$ TeV. This is helped by the
robust prediction of the charged and neutral wino mass difference, $\delta m=165-190$~MeV.

\noindent
4)We have also estimated the monochromatic $\gamma$-ray signal coming from the pair
annihilation of wino DM at the galactic centre.  One finds a viable signal upto wino masses of
$2.3$ TeV for cuspy DM density profiles.  
Inclusion of non-perturbative effects would increase this limit by about 
25$\%$. 

\noindent
{\bf Acknowledgments}\\
We thank Manuel Drees, 
Gian Giudice, Antonio Masiero and Stefano Profumo for many
helpful discussions.  DPR acknowledges 
the hospitality of CERN Theory 
Division, where this work was initiated and receiving  
partial financial support 
from BRNS (DAE) under the Raja Ramanna Fellowship Scheme.  
DD would like to thank the Council of Scientific and 
Industrial Research, Govt. of India for the support received as a
Senior Research Fellow.

\end{document}